# VON NEUMANN ENTANGLEMENT AND DECOHERENCE IN TWO DIMENSIONAL QUANTUM WALKS

**CLEMENT AMPADU**

31 Carrolton Road
Boston, Massachusetts, 02132
USA
e-mail: drampadu@hotmail.com

## Abstract

Using the concept of von Neumann entropy, we quantify the information content of the various components of the quantum walk system, including the mutual information between its subsystems (coin and position) and use it to give a precise formulation of the measure of entanglement between subsystems.

## I. Introduction

Quantum entanglement plays a key role in quantum mechanics in view of its connection with the quantum walk. The studies in this area are very extensive and include the effect of entanglement of the coin subspace on the quantum walk [1-3], entanglement between the coin and position subspaces [4-6] and entanglement generation in quantum walks[7-8]. Other studies on quantum entanglement from a numerical and experimental perspective include [4-6,8-9].
Experimentally the physical implementation of the quantum walk faces many obstacles including environmental noise which fades the quantum properties, however the formulation and quantification of the influence of decoherence on the quantum walk has received considerable attention in the literature [10-24]. With respect to mathematical modeling of decoherence and the pure analytic papers, some of the studies include [13-15,17,25-32]. For a recent review of decoherence in quantum walks, the reader is referred to [33].

As Liu and Petulante [34] have pointed out there has not been much effort to give a precise formulation of the relationship between entanglement and decoherence in the literature. In this paper we use the concept of von Neumann entropy to give such a formulation for quantum walks on the square lattice restricted in a sense. The study of the properties of the quantum walk using basic quantities in information theory is not new, and the recent work in this area include [35-37].

This paper is organized as follows. In Section II we define the elements of the quantum random walk on the square lattice restricted to $Z_N \times Z_N$. In Section III we present the main results concerning the long-term trend of the entropies associated with the reduced density operators of coin and position subsystems and the mutual information between them. Section IV is devoted to the conclusions.

## II. Quantum walks on the square lattice subject to decoherence

In the two dimensional quantum walk the "coin" degree of freedom is represented by two-quibit space or coin space, $H_C$, which is spanned by four orthonormal states $\{|j\rangle, j = 1,2,3,4\}$. The position space of the walker, $H_P$, is spanned by the set of orthonormal states $\{|x, y\rangle : x, y \in Z_N\}$. The Hilbert space of the walker is $H = H_P \otimes H_C$. To define the movement of the quantum walker on the square lattice, we first consider what happens on one step in the quantum walk. We first make superposition on the coin space with coin operator $U_C$ and make a move according to the coin state with translation operator $S$ as follows $U = S(I \otimes U_C)$, where $I$ is the identity operator in $H_P$, $U$ is the coin operator on the position space, and the translation operator is given by

$$S = \sum_{x,y} \{|x-1, y\rangle\langle x, y| \otimes |L\rangle\langle L| + |x+1, y\rangle\langle x, y| \otimes |R\rangle\langle R| + |x, y-1\rangle\langle x, y| \otimes |D\rangle\langle D| + |x, y+1\rangle\langle x, y| \otimes |U\rangle\langle U|\}.$$

In terms of the initial state, the evolution of the walk is given by $\psi_{t+1} = U^t \psi_0$. For $\psi \in H$, we can write $\psi = \sum_{x,y}\sum_{j}\psi(x,y,j)|x,y\rangle \otimes |j\rangle$. For the quantum walk with initial state $\psi_0$, the probability $P(x,y,t)$ of finding the walker at $(x,y) \in Z_N \times Z_N$ at time $t$ is given by

$P(x,y,t) = Tr(|x,y\rangle\langle x,y|\rho(t))$, where the time-dependent density operator $\rho(t)$ is defined by

$\rho(t) = \psi_t{}^T\overline{\psi}_t = U^t|\psi_0\rangle\langle\psi_0|({}^T\overline{U})^t$. Let $\{A_n\}_{0 \le n \le v}$ be a unital family of operators satisfying

$\sum_{0 \le n \le v}({}^T\overline{\hat{A}})\hat{A}_n = I$. In the evolution of the walk, for events induced by decoherence we apply to the coin degree of freedom, at each time step of the walk the unital family of operators. In terms of the unital family of operators we can write the density operator as $\rho(t+1) = \sum_{0 \le n \le v} U\hat{A}_n \rho(t)({}^T\overline{\hat{A}})({}^T\overline{U})$.

Let us assume the quantum walk is launched from the position $|0,0\rangle$ in coin state $\psi_0$. We should remark then that we can write the density operator as

$\rho(t) = \frac{1}{N^2}\sum_{k_x,k_y,k'_x,k'_y}|k_x,k_y\rangle\langle k'_x,k'_y| \otimes L^t_{k_x,k_y,k'_x,k'_y}|\psi_0\rangle\langle\psi_0|$, where

$|k_x,k_y\rangle = \frac{1}{N}\sum_{x,y}\exp(2\pi i(xk_x + yk_y)/N)|x,y\rangle$. Here $L_{k_x,k_y,k'_x,k'_y}$ is defined by

$L_{k_x,k_y,k'_x,k'_y}|\psi_0\rangle\langle\psi_0| = \sum_n U_c(k_x,k_y)\hat{A}_n|\psi_0\rangle\langle\psi_0|{}^T\overline{\hat{A}}^T\overline{U(k'_x,k'_y)} = (B_{ij}(1,k_x,k_y,k'_x,k'_y))_{1 \le i,j \le 4}$, where

$U_C(k_x,k_y) = Diag\left(e^{\frac{-2\pi i k_x}{N}}, e^{\frac{2\pi i k_x}{N}}, e^{\frac{-2\pi i k_y}{N}}, e^{\frac{2\pi i k_y}{N}}\right)U_C$. After $t$ iterations we can write

$L^t_{k_x,k_y,k'_x,k'_y}|\psi_0\rangle\langle\psi_0| = (B_{ij}(t,k_x,k_y,k'_x,k'_y))_{1 \le i,j \le 4}$. In terms of the $4N^2$-dimensional Hilbert space $H = H_P \otimes H_C$, the density operator is given by $\rho(t) = \sum_{1 \le x,y,z,v \le N}\sum_{1 \le l,m \le 2}P_{xyzvlm}(t)|x,y\rangle\langle z,v| \otimes |l\rangle\langle m|$, where $P_{xyzvlm}(t) = \frac{1}{N}\sum_{k_x,k_y,k'_x,k'_y}\langle x,y|k_x,k_y\rangle\langle k'_x,k'_y|z,v\rangle B_{lm}(t,k_x,k_y,k'_x,k'_y)$. We should note that

the probability $P(x,y,t)$ of finding the walker at $(x,y)$ at time $t$ is given by

$$P(x, y, t) = Tr(|x, y\rangle\langle x, y|\rho(t)) = \frac{1}{N} \sum_{k_x, k_y} \sum_{k'_x, k'_y} \langle x, y|k_x, k_y\rangle\langle k'_x, k'_y|x, y\rangle Tr\left(L^t_{k_x k_y k'_x k'_y} |\psi_0\rangle\langle\psi_0|\right)$$

$$= \frac{1}{N} \sum_{k_x, k_y, k'_x, k'_y} \langle x, y|k_x, k_y\rangle\langle k'_x, k'_y|x, y\rangle Tr\left(L^t_{k_x k_y k'_x k'_y} |\psi_0\rangle\langle\psi_0|\right)$$

$$= \frac{1}{N} \sum_{k_x, k_y, k'_x, k'_y} \langle x, y|k_x, k_y\rangle\langle k'_x, k'_y|x, y\rangle \left(\sum_l B_{ll}(t, k_x, k_y, k'_x, k'_y)\right)$$

$$= P_{xyxy11}(t) + P_{xyxy22}(t) + P_{xyxy33}(t) + P_{xyxy44}(t)$$

In this paper we will take $U_C = \frac{1}{2}\begin{bmatrix} 1 & 1 & 1 & 1 \\ 1 & -1 & 1 & -1 \\ 1 & 1 & -1 & -1 \\ 1 & -1 & -1 & 1 \end{bmatrix}$, which is coin operator for the standard two dimensional Hadamard walk. We should note that in the Fourier picture the coin operator takes the form $U_C(k_x, k_y) = \frac{1}{2}\begin{bmatrix} e^{\frac{-2\pi i k_x}{N}} & e^{\frac{-2\pi i k_x}{N}} & e^{\frac{-2\pi i k_x}{N}} & e^{\frac{-2\pi i k_x}{N}} \\ e^{\frac{2\pi i k_x}{N}} & -e^{\frac{2\pi i k_x}{N}} & e^{\frac{2\pi i k_x}{N}} & -e^{\frac{2\pi i k_x}{N}} \\ e^{\frac{-2\pi i k_y}{N}} & e^{\frac{-2\pi i k_y}{N}} & -e^{\frac{-2\pi i k_y}{N}} & -e^{\frac{-2\pi i k_y}{N}} \\ e^{\frac{2\pi i k_y}{N}} & -e^{\frac{2\pi i k_y}{N}} & -e^{\frac{2\pi i k_y}{N}} & e^{\frac{2\pi i k_y}{N}} \end{bmatrix}$. In this paper we will take the unital family $\{\hat{A}_n\}_{0 \le n \le v}$ of operators as $\hat{A}_0 = Diag(1-p, 1-p, 1-p, 1-p)$, $\hat{A}_1 = Diag(p, 0, 0, 0)$,

$\hat{A}_2 = Diag(0, 0, 0, p)$, $\hat{A}_3 = Diag(0, p, 0, 0)$, $\hat{A}_4 = Diag(0, 0, p, 0)$, where $0 \le p \le 1$, and $p$ is

the decoherence rate. The value of $p$ plays a significant role in the evolution of the quantum walk subject to decoherence. If $p = 0$, the quantum walk evolves as a purely quantum process, and when $p = 1$ behaves exactly like the classical random walk in two dimensions, in which the probability of moving in each of the axial directions is $\frac{1}{4}$. Now let $L(C^4)$ denote the Hilbert space of all $4 \times 4$ matrices with the inner product given by $\langle M_1, M_2\rangle \equiv tr(^T\overline{M}_1 M_2)$. Now we list the

necessary tools essential to our analysis, the proofs are similar in nature to those in Liu and

Pentulante [34], upon making the necessary change in the dimension of the Hilbert space, therefore we omit them except special cases.

**Lemma 1:** Let S be a superoperator on the Hilbert space $L(C^4)$ defined by

$$S = \sum_{n=0}^{4} U_1 \hat{A}_n^T \overline{\hat{A}}_n U_2 : B \mapsto \sum_{n=0}^{4} U_1 \hat{A}_n B^T \overline{\hat{A}}_n U \text{ , where } U_1, U_2 \text{ are } 4 \times 4 \text{ unitary matrices and}$$

$B \in L(C^4)$. Then $\langle SB, SB \rangle \leq \langle B, B \rangle$. In particular, for all $B \in L(C^4)$, $\langle SB, SB \rangle = \langle B, B \rangle$ iff $p = 0$.

As a consequence of Lemma 1 we can get the following

**Corollary:** $|\lambda| \leq 1$ for every eigenvalue $\lambda$ of $S$.

**Proof:** Let $B_\lambda$ be an eigenvector of $S$ belonging to $\lambda$, then $\langle SB_\lambda, SB_\lambda \rangle \leq \langle B_\lambda, B_\lambda \rangle = |\lambda|^2 \langle B_\lambda, B_\lambda \rangle$. Applying Lemma 1 gives the desired result.

Now consider the superoperator $L_{k_x,k_y,k'_x,k'_y} : L(C^4) \mapsto L(C^4)$, let a basis for $L(C^4)$ be given by

the following 
$$\begin{cases} \sigma_o \otimes \sigma_o, \sigma_o \otimes \sigma_x, \sigma_o \otimes \sigma_y, \sigma_o \otimes \sigma_z, \sigma_x \otimes \sigma_y, \sigma_x \otimes \sigma_z, \sigma_y \otimes \sigma_z, \sigma_x \otimes \sigma_o, \\ \sigma_y \otimes \sigma_o, \sigma_z \otimes \sigma_o, \sigma_y \otimes \sigma_x, \sigma_z \otimes \sigma_x, \sigma_z \otimes \sigma_y, \sigma_y \otimes \sigma_y, \sigma_z \otimes \sigma_z, \sigma_x \otimes \sigma_x \end{cases},$$

where $\sigma_o$, $\sigma_x$, $\sigma_y$, and $\sigma_z$ are the Pauli matrices, then in terms of the basis, a matrix representation of $L_{k_x k_y k'_x k'_y}$ is given by the following, but because of the dimension of the matrix, and hence its huge size, we will leave the matrix representation in tensor form as follows :

$$\begin{bmatrix} c^- & iqs^- & 0 & 0 \\ 0 & 0 & qs^+ & c^+ \\ 0 & 0 & -qc^+ & s^+ \\ is^- & qc^- & 0 & 0 \end{bmatrix} \otimes \begin{bmatrix} c^- & iqs^- & 0 & 0 \\ 0 & 0 & qs^+ & c^+ \\ 0 & 0 & -qc^+ & s^+ \\ is^- & qc^- & 0 & 0 \end{bmatrix}, \text{ where } s^\pm = \sin \frac{2\pi(k \pm k')}{N} \text{ and }$$

$c^\pm = \cos \frac{2\pi(k \pm k')}{N}$, and we have let $q = 1 - p$. To write the resulting matrix in terms of

$k_x, k_y, k'_x, k'_y$, write the matrices $\left(B_{ij}(1, k_x, k_y, k'_x, k'_y)\right)_{1 \leq i,j \leq 4}$ as given above and $\left(B_{ij}(1, k, k')\right)_{1 \leq i,j \leq 2}$ defined in a similar way, in terms of their *basis*, and compare the matrix resulting from $\left(B_{ij}\left(1, k_x, k_y, k'_x, k'_y\right)\right)_{1\leq i, j\leq 4}$ with the tensor product of the matrix resulting from $\left(B_{ij}(1, k, k')\right)_{1\leq i, j\leq 2}$ with itself.

It can be shown that the characteristic polynomial associated with the matrix

$$\begin{bmatrix} c^- & iqs^- & 0 & 0 \\ 0 & 0 & qs^+ & c^+ \\ 0 & 0 & -qc^+ & s^+ \\ is^- & qc^- & 0 & 0 \end{bmatrix}$$ is given by

$f(\lambda) = \lambda^4 + (qc^+ - c^-)\lambda^3 - 2qc^+c^-\lambda^2 + q(c^+ - qc^-)\lambda + q^2$, where $q$, $c^{\pm}$ are defined as above.

It follows that the characteristic polynomial of $L_{k_x k_y k'_x k'_y}$ is given by $g(\lambda) = [f(\lambda)]^4$. It should be noted that $g(\lambda) = [f(\lambda)]^4$ is given in terms of $k, k'$, therefore to write it in terms of $k_x, k'_x, k_y, k'_y$, the equivalence between the matrix $\left(B_{ij}\left(1, k_x, k_y, k'_x, k'_y\right)\right)_{1\leq i, j\leq 4}$ in terms of its basis, and the tensor product of the matrix $\left(B_{ij}(1, k, k')\right)_{1\leq i, j\leq 2}$ in terms of it basis, can be used. It follows from Liu and Pentulante [34], that we have the following result concerning the eigenvalues of $L_{k_x k_y k'_x k'_y}$.

**Proposition 1:** Suppose $0 < p < 1$. Let a typical eigenvalue of $L_{k_x k_y k'_x k'_y}$ be denoted by $\lambda$. Then:

(i) $\|\lambda\| < 1$

(ii) If $\|\lambda\| = 1$, then $\lambda = \pm 1$

(iii) $\lambda = 1$, iff $k_x = k'_x$ and $k_y = k'_y$, in which case $\lambda = 1$ has algebraic multiplicity 1

(iv) $\lambda = -1$ iff $|k_x - k'_x| = \dfrac{N}{4}$ and $|k_y - k'_y| = \dfrac{N}{4}$, in which case $\lambda = -1$ has algebraic multiplicity 1

We should remark that Proposition 1 allows us to specify the long-term behavior of the matrix components of the total density operator $\rho(t) = \sum\limits_{1\leq x, y, z, v\leq N} \sum\limits_{1\leq l, m\leq 2} P_{xyzvlm}(t)|x, y\rangle\langle z, v| \otimes |l\rangle\langle m|$. The starting point is the following.

**Proposition 2:** For the matrix in $L^t_{k_x k_y k'_x k'}|\psi_0\rangle\langle\psi_0| = \left(B_{ij}\left(t, k_x, k_y, k'_x, k'_y\right)\right)_{1\leq i, j\leq 4}$, we have the following:

(i) Suppose $k_x = k'_x$ and $k_y = k'_y$. If $i = j$, then $\lim_{t \to \infty} B_{ij}(t, k_x, k_y, k'_x, k'_y) = \frac{1}{4}$. If $i \neq j$, then

$$\lim_{t \to \infty} B_{ij}(t, k_x, k_y, k'_x, k'_y) = 0.$$

(ii) Suppose $|k_x - k'_x| = \frac{N}{4}$, $|k_y - k'_y| = \frac{N}{4}$. If $i = j$, then $\lim_{t \to \infty} (-1)^t B_{ij}(t, k_x, k_y, k'_x, k'_y) = \frac{1}{4}$.

If $i \neq j$, then $\lim_{t \to \infty} B_{ij}(t, k_x, k_y, k'_x, k'_y) = 0$.

(iii) Suppose $|k_x - k'_x| \neq \frac{N}{4}, 0$ and $|k_y - k'_y| \neq \frac{N}{4}, 0$. Then for all combinations of $i, j$, we have

$$\lim_{t \to \infty} B_{ij}(t, k_x, k_y, k'_x, k'_y) = 0.$$

From Proposition 2, we get the long time behavior of the operator in

$$L^t_{k_x k_y k'_x k'_y} |\psi_0\rangle\langle\psi_0| = \left(B_{ij}(t, k_x, k_y, k'_x, k'_y)\right)_{1 \leq i, j \leq 4} \text{ as follows.}$$

**Theorem 3:** Consider the quantum on the square lattice, more specifically, $Z_N \times Z_N$, let the total density operator be as given previously.

(i) Suppose $N$ is odd. If $x = z$, $y = v$ and $l = m$, then $\lim_{t \to \infty} P_{xyzvlm}(t) = \frac{1}{4N}$. If

$x \neq z$, $y \neq v$, or $l \neq m$, then $\lim_{t \to \infty} P_{xyzvlm}(t) = 0$

(ii) Suppose $N$ is even. If $x = z$, $y = v$ and $l = m$, and if $t - x$ and $t - y$ are both even, then

$\lim_{t \to \infty} P_{xyzvlm}(t) = \frac{1}{N}$. Otherwise, $\lim_{t \to \infty} P_{xyzvlm}(t) = 0$.

### III. On the relationship between entanglement and decoherence

The purpose of this section is to answer the following question: Under what conditions on the decoherence rate $p$, must the mutual information between the subsystem of the coin and the subsystem of the walker eventually diminish to zero?

We should remark that for quantum walks on the $N-$cycle, Liu and Petulante [34] have shown that the answer is $p > 0$. Before we commence the investigation, some necessary definitions and remarks on the von Neumann entropy of a quantum system and mutual information are necessary.

**Definition 4 (von Neumann Entropy of a Quantum System):** The von Neumann Entropy of a quantum system $A$, denoted $S(A)$, is a measure of the uncertainty implied by the multitude of potential outcomes as reflected by its density matrix $\rho(A)$. By definition, $S(A) = S(\rho(A)) = -Tr(\rho \ln \rho)$. For composite systems $A$ and $B$, the von Neumann entropy of the composite system, $S(A,B)$, is defined by $S(A,B) = -Tr(\rho^{AB} \ln \rho^{AB})$, where $\rho^{AB}$ is the density matrix of the composite system.

We should remark that in a quantum walk with zero decoherence the entropy of the reduced density operator on the coin subsystem can serve as measure of its degree of entanglement relative to the subsystem of the walker, whilst to measure the level of quantum entanglement in a quantum walk subject to non-zero decoherence, the von Neumann entropy must be considered separately for each of the subsystems as well for the total system.

**Definition 5 (Mutual Information):** Given two systems $A, B$, their mutual information, $S(A:B)$, is defined by the formula, $S(A:B) = S(A) + S(B) - S(A,B)$.

We should remark that the level of entanglement between the systems $A, B$ is contained in the mutual information.

Now to answer the question posed at the beginning of this section, we begin with the following from Liu and Pentulante [34] whose proof is due to Watrous [38].

**Lemma 6:** Let $X$ denote a complex Euclidean space and let $Pos(X)$ denote the set of all positive semi-definite operators defined on $X$ with the norm $\|\rho\|_{tr} = Tr\left(\sqrt{\overline{\rho}^T \rho}\right)$. Then, with respect to this norm, the von Neumann entropy $S(\rho)$ is continuous at every point $\rho \in Pos(X)$.

If we let $\rho_{N_{ODD}}(\infty) = diag\left(\frac{1}{4N}, \cdots, \frac{1}{4N}\right)$, then Theorem 3 implies $\lim_{t \to \infty} \|\rho(t) - \rho_{N_{ODD}}(\infty)\|_{tr} = 0$. So by Lemma 6 $\lim_{t \to \infty} S(\rho(t)) = \lim_{t \to \infty} S(\rho_{N_{ODD}}(\infty)) = 1 + \ln N$. In the case ,

$\rho_{N_{EVEN}}(\infty) = diag\left(\dfrac{1}{N},\cdots,\dfrac{1}{N}\right)$, we also have by Theorem 3, $\lim_{t\to\infty}\|\rho(t) - \rho_{N_{EVEN}}(\infty)\|_{tr} = 0$. So by

Lemma 6 $\lim_{t\to\infty} S(\rho(t)) = \lim_{t\to\infty} S(\rho_{N_{EVEN}}(\infty)) = \ln N$. In particular we have the following

**Theorem 7:** Suppose the quantum walk is launched on the square lattice, specifically $Z_N \times Z_N$ with initial coin state $|\psi_0\rangle$ and decoherence rate $\rho > 0$, let $\rho(t)$ be the density operator of the overall system, then, $\lim_{t\to\infty} S(\rho(t)) = \begin{cases} 1+\log N, \text{if } N \text{ is odd} \\ \log N, \text{if } N \text{ is even} \end{cases}$.

Now let $\rho_C(t) = tr_W(\rho(t))$ be the time dependent reduced density operator for the subsystem associated with the coin and let $\rho_W(t) = tr_C(\rho(t))$ be the time dependent reduced density operator for the subsystem associated with the walker. Combining Theorem 7 with the inequality $S(\rho_C(t), \rho_W(t)) \leq S(\rho_C(t)) + S(\rho_W(t)) \leq 1 + \ln N$, implies $\lim_{t\to\infty} S(\rho_C(t), \rho_W(t)) = 1 + \ln N$, thus, $\lim_{t\to\infty}[S(\rho_C(t)) + S(\rho_W(t))] = 1 + \ln N$, thus from Definition 5 we see that $\lim_{t\to\infty} S(\rho_C(t) : \rho_W(t)) = 0$. In particular we have the main result.

**Theorem 8:** Suppose the quantum walk is launched on the square lattice, specifically $Z_N \times Z_N$ with initial coin state $|\psi_0\rangle$ and decoherence rate $\rho > 0$, then $\lim_{t\to\infty} S(\rho_C(t) : \rho_W(t)) = 0$.

IV. **Concluding remarks**

In this paper we have shown that whenever the decoherence rate is $p > 0$, the mutual information between the subsystem of the coin and the subsystem of the walker eventually diminish to zero by way of Theorem 8. We have also shown by way of Theorem 3 that decoherence in the coin-driven quantum walk on the square lattice, precisely $Z_N \times Z_N$, is characterized by the vanishing of the off-diagonal elements in the density matrix which present the quantum correlations.